\documentstyle[12pt]{article}

\def\simge{\mathrel{%
   \rlap{\raise 0.511ex \hbox{$>$}}{\lower 0.511ex \hbox{$\sim$}}}}
\def\simle{\mathrel{
   \rlap{\raise 0.511ex \hbox{$<$}}{\lower 0.511ex \hbox{$\sim$}}}}
 
\def\slashchar#1{\setbox0=\hbox{$#1$}           
   \dimen0=\wd0                                 
   \setbox1=\hbox{/} \dimen1=\wd1               
   \ifdim\dimen0>\dimen1                        
      \rlap{\hbox to \dimen0{\hfil/\hfil}}      
      #1                                        
   \else                                        
      \rlap{\hbox to \dimen1{\hfil$#1$\hfil}}   
      /                                         
   \fi}                                         %

\def\ts{\thinspace}

\def\ra{\rightarrow}

\def\ol{\bar}

\def\be{\begin{equation}} 
\def\ee{\end{equation}} 
\def\bea{\begin{eqnarray}}
\def\eea{\end{eqnarray}}
\def\ba{\begin{array}}
\def\ea{\end{array}}

\def\tev{{\rm TeV}}

\def\half{{\textstyle{ { 1\over { 2 } }}}}

\begin{document}
\title{
\vskip -15mm
\begin{flushright}
\vskip -15mm
{
\small BUHEP-01-3\\
\small hep-ph/0102131\\}
\vskip 5mm
\end{flushright}
{\Large{\bf \hskip 0.38truein
New Model--Independent Limit \\ on Muon Substructure}}\\
}
\author{
\centerline{{Kenneth Lane\thanks{lane@physics.bu.edu}}}\\ \\
\centerline{{Department of Physics, Boston University,}}\\
\centerline{{590 Commonwealth Avenue, Boston, MA 02215}}\\
}
\maketitle
\begin{abstract}
If the discrepancy between the theoretical and newly measured values of the
muon's anomalous magnetic moment is ascribed to muon substructure, there
results an improved model--independent limit on its energy scale, $1.2\,\tev
< \Lambda_\mu < 3.2\,\tev$ at 95\%~C.L.
\end{abstract}


\newpage

The Muon ($g-2$) Collaboration has announced a new measurement of the
anomalous magnetic moment of the positive muon~\cite{gtwo},
\be\label{eq:meas} 
a_{\mu^+} = \half(g-2)_{\mu^+} = 11\,\,659\,\,202(14)(6) \times 10^{-10}
\ts\ts (1.3\ts {\rm ppm}) \ts.
\ee
The value currently expected in the standard model is~\cite{smvalue}
\be\label{eq:SM}
a_{\mu}({\rm {SM}}) = 11\,\,659\,\,159.6(6.7)\times 10^{-10}
\ts\ts (0.57\ts {\rm ppm})\ts.
\ee
Using the world--average experimental value of the muon's anomalous moment,
there is now a discrepancy with theory of 2.6 standard deviations:
\be\label{eq:diff}
\delta a_\mu \equiv a_\mu(\rm exp) - a_{\mu}({\rm {SM}}) = 43(16) \times
10^{-10}\ts.
\ee
This error is smaller than that of previous measurements by about a factor of
three~\cite{previous}.

If the muon is a composite fermion at the scale $\Lambda_\mu \gg m_\mu$,
there is a contribution to its magnetic moment of~\cite{composite}
\be\label{eq:delta}
\delta a_\mu(\Lambda_\mu) \simeq {m_\mu^2 \over{\Lambda_\mu^2}} \ts.
\ee
If $\delta a_\mu$ is ascribed to muon substructure, the scale is $\Lambda_\mu
\simeq 1.6\,\tev$. It is more appropriate, however, to take advantage of the
new smaller error by interpreting $\delta a_\mu$ to limit $\Lambda_\mu$. For
example, its 95\%~C.L. range is
\be\label{eq:limit}
1.2\,\tev < \Lambda_\mu < 3.2\,\tev \ts.
\ee
If, as expected, analysis of the 2000 data for $g-2$ decreases its
statistical error by half, but the central value and other errors do not
change, this range will become $1.3\,\tev < \Lambda_\mu < 2.3\,\tev$

The importance of this bound is its model--independence. It requires no
assumption on the compositeness of other quarks or leptons. Limits that do
assume $\Lambda_e \simeq \Lambda_\mu$ and $\Lambda_{u,d} \simeq \Lambda_\mu$
come from $e^+ e^- \ra \mu^+\mu^-$ and $\ol qq \ra
\mu^+\mu^-$~\cite{elp}. They are $\Lambda_e \simeq \Lambda_\mu \simge
4$--$5\,\tev$ and $\Lambda_{u,d} \simeq \Lambda_\mu \simge
3$--$4\,\tev$~\cite{pdg}. These lower bounds are more stringent, but they are
also less incisive. There is no reason {\it a priori} for the equality of the
first and second--generation lepton substructure scales.

\section*{Acknowledgements}

I thank Robert Carey for a discussion of the new $g-2$ result of
Ref.~\cite{gtwo} and Estia Eichten, Bill Marciano, Jim Miller, and Lee
Roberts for valuable comments. This research was supported in part by the
Department of Energy under Grant~No.~DE--FG02--91ER40676.

\vfil\eject

\end{document}